\documentclass[aps,prl,preprint]{revtex4-2}

\usepackage{amssymb,amsmath,amsfonts}
\usepackage{graphicx,epstopdf}

\begin{document}

\title{Self-action of electromagnetic charge in a wormhole with an infinitely short throat}
\author{O. Aslan}
\affiliation{Kazan Federal University, 420008, Kremlyovskaya street 18, Kazan,  Russia}
\email{alsucuk@gmail.com}

\author{A. Popov}
\affiliation{Kazan Federal University, 420008, Kremlyovskaya street 18, Kazan, Russia}
\email{apopov@kpfu.ru}

\begin{abstract}{The self-force of the electromagnetic charge in the space-time of a wormhole with an infinitely short throat is calculated. It is assumed that the charge is a source of an electromagnetic field that is in non-minimal connection with the curvature of space-time.}
\end{abstract}



\maketitle


\section{Introduction}

A charged particle at rest in a curved space-time generates a field that, due to the curvature of space-time and the non-local structure of the massless field, affects the particle itself. Such a force is called the self-force \cite{Khusnutdinov:2021, PoissonPoundVega:2011}
 A similar situation occurs in the case of gravitational charge \cite{Oltean:2019, ZimmermanPoisson:2014, BiniGeralico1:2019}.

Electromagnetic and gravitational forces of self-interaction are important when describing the motion of two bodies with an extreme mass ratio and when studying the gravitational waves emitted by them \cite{Brack:2009, BrackPound:2018}.
In curved spaces, intensive studies of the self-interaction of a stationary charge were conducted on the background of black holes, spaces of topological defects, and wormholes.

In cases of static charges in the Schwarzschild space, the electrostatic potential and electromagnetic self-force are known explicitly \cite{Copson:1928, HanniRuffini:1973, Smith:1980, Lohiya:1982}. In the Reissner-Nordstrom black hole, the electrostatic potential was obtained in \cite{LeauteLinet:1976, Zel-Frol:1982}, and in Kerr's spacetime in \cite{Linet:1982, Leaute:1977}. The self-force of a scalar charge at rest in a Kerr-Newman black hole was considered in \cite{BurkoLiu:01}.

The studies in spaces of topological defects (for example, an infinitely long straight cosmic string, a global monopole) help understand the self-interaction better, since this effect is sensitive not only to curvature, but also to the topological structure of space-time \cite{BezerraSaharian:2007, Linet:1986, Linet2:1986, Smith:1986}

Wormhole spaces are also interesting for studying the self-interaction, since they have both a non-trivial topological structure and curvature.
For static scalar and electromagnetic charges, the effect of self-interaction in space-time of wormholes with different throat shapes were considered in \cite{BezKhu:09, Popov:13, Taylor:13}. It is interesting to note that in the case of a long throat, the self-interaction is local even for a massless scalar field, nonminimally coupled to the curvature of spacetime \cite{PopovAslan:2015}.

The purpose of this paper is to analyze the self-interaction for a charge in the space-time of a wormhole with an infinitely short throat. It is assumed that the charge is the source of the electromagnetic field that is nonminimally coupled to the curvature of space-time.
The article is organized as follows: In section II, we obtain an unnormalized expression for the potential of self-interaction of a static electromagnetic charge on the gravitational background under consideration. Section III describes the procedure of renormalization of the potential of self-interaction and the result.

In this work we use the following definitions of the curvature tensor $R_{ABC}^D=\partial_C\Gamma_{AB}^D-\partial_B\Gamma_{AC}^D+\Gamma_{EC}^D\Gamma_{BA}^E-\Gamma_{EB}^D\Gamma_{AC}^E$ and
the Riemann tensor $R_{MN}=R^F_{MFN}$.
The system of units is chosen so that $c=G=1$.

\section{Self-force computations}

Consider a static spherical symmetrical wormhole
\begin{equation}\label{metric}
ds^2 = -dt^2 + dr^2 + L(r)^2\left( d\theta^2 +\sin^2 \theta d \varphi^2 \right),
\end{equation}
where  $-\infty < r < \infty, \ \theta \in [0, \pi], \ \varphi \in [0, 2\pi)$. \\
The simplest model of a wormhole is a model with an infinitely short throat, which corresponds to
\begin{equation} \label{L}
L(r)=(|r|+a),\ a>0,
\end{equation}
which we will consider in further. Such a model represents two Minkowski spacetimes in each of which a ball of radius $a$ is cut out and glued along the surface of these balls.
As known, this model works well when describing effects at large (compared to the length of the throat) distances from the throat of the wormhole.

The equations of the field created by an electromagnetic charge can be obtained by varying the potential of the electromagnetic field of the corresponding action. We will consider Einstein's theory of gravity with a linear by curvature coupling between the electromagnetic and gravitational fields, for which the action has the form
\cite{Balakin:2005}
\begin{equation}
S = \int d^4 x \sqrt{-g} \, {\cal L} = \frac{1}{16 \pi} \int d^4 x \sqrt{-g} \, \left(R -  F_{mn}F^{mn}
+ \,{\chi}^{ikmn} F_{ik}F_{mn}\right) \,,
\label{simplifiednonminimal}
\end{equation}
where $g$ - the determinant of the metric tensor $g_{i k}$, $R$ is the scalar curvature of space-time, the value
\begin{equation}
{\chi}^{ikmn} \equiv  \frac{q_1 R}{2}(g^{im}g^{kn} {-} g^{in}g^{km}) {+}
\frac{q_2}{2} (R^{im}g^{kn} {-} R^{in}g^{km} {+} R^{kn}g^{im} {-}
R^{km}g^{in}) {+} q_3 R^{ikmn}
\label{susceptibilitytensor1}
\end{equation}
is called the susceptibility tensor,
\begin{equation}
F_{ik} = \nabla_i A_{k} - \nabla_k A_{i} = \frac{\partial
A_{k}}{\partial x^i} - \frac{\partial A_{i}}{\partial x^k} ,  \label{6}
\end{equation}
is the electromagnetic field tensor, $R^{in}$ is the Ricci tensor, $R^{ikmn}$ is the tensor of curvature. The parameters $q_1, q_2$, and $q_3$ are in general arbitrary.

Variation of the potential, $A_k$ of the action(\ref{simplifiednonminimal}) gives the electromagnetic field equations
\begin{equation}
\nabla_k H^{ik} = 0 \,,
\label{nonm}
\end{equation}
where $H^{ik}$ is the induction tensor defined by the expression
\begin{equation}
H^{ik} \equiv F^{ik} - {\chi}^{ikmn} F_{mn}  \,.
\label{inductiontensor}
\end{equation}
If the field $A_i(x^k)$ is created by the charge  $e$ , the equations (\ref{nonm}) are converted to the form
\begin{equation}
\nabla^k H_{ik} = - 4 \pi j_i= - 4 \pi  e \int u_i(\tau)
\delta^{(4)}(x^k,\tilde x^k(\tau)) \frac{d\tau}{ \sqrt{-g}},
\label{feq}
\end{equation}
where $u^i$ - 4-speed of the charge and $\tau$ is its proper time. The world line of the charge is defined by the functions $\tilde x^k (\tau)$.
For a charge at rest $u^i (1,0,0,0)$, and the vector potential $A_i$ does not
depend on the time, which allows us to use the following ansatz: $A_i = (A_t,0,0,0)$.
The Lorentz calibration in this case is performed identically, and the system of equations
\eqref{feq} reduce to a single equation for $A_t$:

\begin{eqnarray}
&& \left[-1 +\frac{8 \, \delta(r)}{(|r|+a)}q_1 +\frac{4 \, \delta(r)}{(|r|+a)}q_2  \right]\frac{d^2 A_t}{d r^2}
+\left[- \frac{2 \, \mbox{sgn}(r)}{(|r| +a)} +\left(\frac{8 \, \delta'(r)}{(|r| +a)}
\right. \right. \nonumber \\ 
&& \left. \left.
+\frac{16 \, \mbox{sgn}(r) \delta(r)}{(|r| +a)^2} \right) q_1
+\left(\frac{4 \, \delta'(r)}{(|r| +a)} +\frac{4 \, \mbox{sgn}(r) \delta(r)}{(|r| +a)^2} \right) q_2 \right]\frac{d A_t}{d r}
+\left[ - \frac{1}{(|r| +a)^2}
\right. \nonumber \\ 
&& \left. + \frac{8 \delta(r)}{(|r| +a)^3} q_1 + \frac{2 \delta(r)}{(|r| +a)^3} q_2
\right]\frac{d^2 A_t}{d \theta^2}
+ \frac{\cos(\theta)}{\sin(\theta)}
\left[ - \frac{1}{(|r| +a)^2}  +\frac{8 \delta(r)}{(|r| +a)^3} q_1
\right. \nonumber \\ && \left.
+ \frac{2 \delta(r)}{(|r| +a)^3} q_2 \right]\frac{d A_t}{d \theta}
+ \frac{1}{ \sin^2(\theta)}
\left[ - \frac{1}{(|r| +a)^2}  +\frac{8 \delta(r)}{(|r| +a)^3} q_1
\right. \nonumber \\ && \left.
+ \frac{2 \delta(r)}{(|r| +a)^3} q_2 \right]
 \frac{d^2 A_t}{d \varphi^2}
= -\frac{4 \pi e \delta^3(r,\theta,\varphi; \tilde r,\tilde \theta, \tilde \varphi)}{(|r| +a)^2 \sin(\theta)},
\end{eqnarray}
where the prime sign $'$ indicates the derivative with respect to $r$.
Since the problem under consideration is spherically symmetric, it is convenient to decompose the potential by angular harmonics
\begin{equation}\label{potential}
A_t(x, \tilde{x} ) = -4\pi e \sum_{l,m} Y_{lm}(\theta, \varphi)Y^*_{lm}(\tilde{\theta}, \tilde{\varphi}) g_l(r, \tilde r)= -e \sum_{l=0}^\infty
\left(2l+1\right) P_l(\cos\gamma) g_l(r,\tilde r),
\end{equation}
where $Y_{lm}(\theta, \varphi)$ spherical functions, $P_l$ - Legendre polynomial, $\cos \gamma \equiv \cos \theta \cos \tilde \theta
+\sin \theta \sin \tilde \theta \cos(\varphi-\tilde \varphi)$.
Due to the properties of spherical functions, the radial part, $g_{l}(r; \tilde r)$, satisfies the equation
\begin{eqnarray} \label{del}
&&\left[-1 +\frac{8 \, \delta(r)}{(|r|+a)}q_1 +\frac{4 \, \delta(r)}{(|r|+a)}q_2  \right]\frac{d^2 g_l(r,\tilde r)}{d r^2}
+\left[- \frac{2 \, \mbox{sgn}(r)}{(|r| +a)} +\left(\frac{8 \, \delta'(r)}{(|r| +a)}
\right. \right. \nonumber \\ && \left. \left.
+\frac{16 \, \mbox{sgn}(r) \delta(r)}{(|r| +a)^2} \right) q_1
+\left(\frac{4 \, \delta'(r)}{(|r| +a)} +\frac{4 \, \mbox{sgn}(r) \delta(r)}{(|r| +a)^2} \right) q_2
\right]\frac{d g_l(r,\tilde r)}{d r}
\nonumber \\ &&
-l(l+1)\left[ - \frac{1}{(|r| +a)^2}
+ \frac{8 \delta(r)}{(|r| +a)^3} q_1 + \frac{2 \delta(r)}{(|r| +a)^3} q_2
\right] g_l(r,\tilde r)
\nonumber \\ &&
=\frac{ \delta(r; \tilde r)}{(|r| +a)^2 }
\end{eqnarray}
The solution of this equation can be presented in the following form
\begin{equation} \label{radialform}
g_l(r; \tilde r) = \theta(r-\tilde r) p_{n l}(r) q_{n l} (\tilde r) + \theta(\tilde r-r) p_{n l}(\tilde r) q_{n l}(r),
\end{equation}
where the modes $p_{n l} (r)$ and $q_{n l} (r)$ satisfy the corresponding homogeneous equations
\begin{eqnarray} \label{eq2}
&&\frac{d}{d r}\left\{
\left[ -L^2 +q_1 \left( -2 +2{L'}^2 +4 L L'' \right) +2 q_2 L L''\right]
\left\{
   {\begin{array}{l}p_{n l}'(r)\\
   q_{n l}'(r)\end{array}}\right\}
   \right\}
\nonumber \\ &&
-l(l+1)\left[ - 1 + \frac{4 L''}{L} q_1 + \frac{L''}{L} q_2\right]
\left\{
   {\begin{array}{l}p_{n l}(r)\\
   q_{n l}(r)\end{array}}\right\}
=0.
\end{eqnarray}
$p_{n l} (r)$ is chosen as the solution that approaches to zero at $r \to + \infty$ and diverges at $r \to - \infty$, and
$q_{n l} (r)$ is chosen as the solution that approaches to zero at $r \to - \infty$ and diverges at $r \to + \infty$,
that is,
\begin{eqnarray}\label{condlimit}
\lim_{r \to +\infty}p_{n l}(r) &=& 0,\ \lim_{r \to +\infty}q_{n l}(r) = \infty, \nonumber \\
 \lim_{r \to -\infty}p_{n l}(r) &=& \infty,\ \lim_{r \to -\infty}q_{n l}(r) =0.
\end{eqnarray}
Normalization of $g_{l}(r; \tilde r)$ is achieved by integrating \eqref{del} over $r$ from $\tilde r -\varepsilon$
to $\tilde r +\varepsilon$ while $\varepsilon \rightarrow 0$. This leads to a condition on Wronskian

        \begin{equation} \label{wronskian}
        \left[p_{n l}\frac{dq_{n l}}{dr}-
        q_{n l}\frac{dp_{n l}}{dr}\right]=\frac{1}{(|r|+a)^2}.
        \end{equation}
Let's denote the region $r>0$ as ${\cal D}_+$, and the region $r<0$ as ${\cal D}_ -$. In flat regions ${\cal D}_+$ and ${\cal D}_ -$, where $L(r)=\pm r+a$,
$L'(r)=\pm 1$, $R(r)=0$ the equation (\ref{eq2}) will take the form
   \begin{eqnarray} \label{modeeqn} \left\{ {
   \frac{d^2}{dr^2}+\frac{2}{r \pm a} \frac{d}{dr}
   -\frac{l(l+1)}{(r \pm a)^2}   }\right\}\left\{
   {\begin{array}{l}p_{n l}\\
   q_{n l}\end{array}}\right\}=0.
   \end{eqnarray}
Independent solutions of these equations have the form
\begin{eqnarray} \label{Fi}
\phi^\pm_{1}(r) &=& (a \pm r)^l , \quad \phi^\pm_{2}(r) = (a \pm r)^{-l-1}, \quad \mbox{for} \ l \geq 0.
\end{eqnarray}
The asymptotics of these solutions for $l > 0$ have the following properties
\begin{eqnarray} \label{asymptotes}
&&\phi_1^+|_{r\to +\infty} \rightarrow \infty,  \qquad  \phi_1^-|_{r\to -\infty} \rightarrow \infty,
\nonumber \\ &&
\phi_2^+|_{r\to +\infty} \rightarrow 0, \qquad \;\; \phi_2^-|_{r\to -\infty} \rightarrow 0.
\end{eqnarray}
General solutions of the equations\eqref{modeeqn} are linear combinations of independent solutions \eqref{Fi}

\begin{eqnarray}\label{solutions}
p_{n l}(r) &=&\left\{ \begin{array}{lc}
                  \alpha^+_1 \phi^+_1 + \beta^+_1 \phi^+_2, & r >0 \\
                  \alpha^-_1 \phi^-_1 + \beta^-_1 \phi^-_2, &
                  r <0
                \end{array} \right., \nonumber \\
q_{n l}(r) &=&\left\{ \begin{array}{lc}
                  \alpha^+_2 \phi^+_1 + \beta^+_2 \phi^+_2, & r >0 \\
                  \alpha^-_2 \phi^-_1 + \beta^-_2 \phi^-_2, &
                  r <0
                \end{array} \right. ,
\end{eqnarray}
where $\alpha_{1,2}^{\pm}$ and $\beta_{1,2}^{\pm}$ are constants.
Substituting the expressions \eqref{solutions} in \eqref{condlimit} and taking into account \eqref{Fi}, we get
\begin{equation}
\alpha_1^+=0, \quad \alpha_2^-=0.
\end{equation}
Then the solutions (\ref{solutions}) are reduced to the following form
\begin{eqnarray}\label{solutions1}
p_{n l}(r) &=&\left\{ \begin{array}{lc}
                  \beta^+_1 \phi^+_2, & r >0 \\
                  \alpha^-_1 \phi^-_1 + \beta^-_1 \phi^-_2, &
                  r <0
                \end{array} \right., \nonumber \\
q_{n l}(r) &=&\left\{ \begin{array}{lc}
                  \alpha^+_2 \phi^+_1 + \beta^+_2 \phi^+_2, & r >0 \\
                  \beta^-_2 \phi^-_2, & r <0
                \end{array} \right. ,
\end{eqnarray}
Substituting these expressions in \eqref{wronskian}, we get
\begin{eqnarray}\label{wr2}
\frac{1}{(|r|+a)^2}& = & \left\{ \begin{array}{lc}
\alpha_2^+ \beta_1^+ \left( \phi_2^+ \frac{\displaystyle d \phi_1^+ }{\displaystyle d r } - \phi_1^+ \frac{\displaystyle d \phi_2^+ }{\displaystyle d r } \right), & r >0 \\
\alpha_1^- \beta_2^- \left( \phi_1^- \frac{\displaystyle d \phi_2^- }{\displaystyle d r } - \phi_2^- \frac{\displaystyle d \phi_1^- }{\displaystyle d r } \right), & r <0 \\
\end{array} \right.
\end{eqnarray}
Substituting (\ref{Fi}) in these expressions, we get
\begin{eqnarray}\label{wr3}
\frac{1}{(|r|+a)^2}& = & \left\{ \begin{array}{lc}
                   \alpha_2^+ \beta_1^+ \left( \frac {\displaystyle 2l+1}{\displaystyle (r+a)^2} \right), & r >0 \\
                   \alpha_1^- \beta_2^- \left( \frac {\displaystyle 2l+1}{\displaystyle (a-r)^2} \right), & r <0 \\
                \end{array} \right.
\end{eqnarray}
Thus, we have obtained the following restrictions on coefficients
\begin{equation} \label{cond0}
\alpha_2^+ \beta_1^+ = \alpha_1^- \beta_2^-= \frac{1}{2\,l+1}
\end{equation}
To find a solution over the entire space-time, we must impose matching conditions
$p_{nl}$ and $q_{nl}$ at the throat $r=0$. The first compliance condition requires
that the solution must be continuous at $r=0$. That is,
\begin{equation} \label{cond1}
p_{nl}(-0)=p_{nl}(+0), \ q_{nl}(-0)=q_{nl}(+0).
\end{equation}
Which means
\begin{eqnarray}\label{cond1__}
   \displaystyle \left.(\alpha^-_1\phi^-_1+\beta^-_1\phi^-_2)\right|_{r = -0} &=& \left. \beta^+_1 \phi^+_2\right|_{r = 0}, \nonumber \\
\left. (\alpha^+_2 \phi^+_1 + \beta^+_2 \phi^+_2)\right|_{r = 0} &=& \left. \beta^-_2 \phi^-_2\right|_{r = -0}
\end{eqnarray}
or
\begin{eqnarray}\label{cond1-1}
\displaystyle \alpha^-_1 a^l +  \beta^-_1 \frac{1}{a^{l+1}} &=&  \beta^+_1 \frac{1}{a^{l+1}},\nonumber \\
\alpha^+_2 a^l + \beta^+_2 \frac{1}{a^{l+1}} &=& \beta^-_2 \frac{1}{a^{l+1}}.
\end{eqnarray}
To obtain the second condition, we integrate the equation \eqref{eq2}
in the range $ (- \epsilon,\epsilon)$, and then go to
limit $\epsilon \to 0$. This gives a second matching condition
\begin{eqnarray} \label{cond2}
\left. \frac{d p_{nl}}{dr}\right|_{r = -0} &=& \left. \frac{d p_{nl}}{dr}\right|_{r = +0} + \frac{2}{a^3} l(l+1)(4 q_1 +q_2) p_{nl}(+0),
\nonumber \\
\left. \frac{d q_{nl}}{dr}\right|_{r = -0} &=& \left. \frac{d q_{nl}}{dr}\right|_{r = +0} + \frac{2}{a^3} l(l+1)(4 q_1 +q_2) q_{nl}(+0).
\end{eqnarray}
Using (\ref{solutions1}) we get
\begin{eqnarray}
\left.\frac{d p_{nl}}{dr}\right|_{r = -0} &=&-\alpha^-_1 l a^{l-1} + \beta^-_1 (l+1) a^{-l-2},\nonumber \\
\left.\frac{d p_{nl}}{dr}\right|_{r = +0} &=&-\beta^+_1 (l+1) a^{-l-2},\nonumber \\
\left.\frac{d q_{nl}}{dr}\right|_{r = -0} &=&\beta^-_2 (l+1) a^{-l-2},\nonumber \\
\left.\frac{d q_{nl}}{dr}\right|_{r = +0} &=&\alpha^+_2 l a^{l-1} - \beta^+_2 (l+1) a^{-l-2}.
\end{eqnarray}
and finally, by inserting these relations into (\ref{cond2}), we get
\begin{eqnarray} \label{cond2-1}
\beta^-_1 + \beta^+_2 - \frac{l}{l+1}\alpha^-_1 a^{2l+1} &=& 2l(4q_1+q_2) \beta^+_1 a^{-2},\nonumber \\
(\beta^-_2 + \beta^+_2) a^{-l+1}- \alpha^+_2\frac{l}{l+1} a^{l+2} &=& 2l(4q_1 + q_2)\alpha^+_2 a^{l} + 2l(4q_1 + q_2)\beta^+_2 a^{-l-1}\, .
\end{eqnarray}
Finding $\alpha^-_1, \alpha^+_2, \beta^-_1, \beta^-_2$ from (\ref{cond1-1}, \ref{cond2-1}) and substituting in the first expression (\ref{cond0})
we get
\begin{eqnarray} \label{b1b2}
\beta^+_1 \beta^+_2=-\frac{a^{2l+1}\Big(a^2-2l(l+1)(4q_1+q_2)\Big)}{2(2l+1)(l+1)\Big(a^2-l(4q_1+q_2)\Big)}.
\end{eqnarray}
If $r > \tilde{r} >0$, one can obtain $g_l (r, \tilde{r})$ using (\ref{radialform}, \ref{solutions1}, \ref{Fi}, \ref{cond0}, \ref{b1b2})
in the following form
\begin{eqnarray} \label{gl}
g_l (r, \tilde{r}) &=& \alpha^+_2 \beta^+_1 \ \phi^+_2 (r)  \phi^+_1 (\tilde{r}) +\beta^+_1 \beta^+_2 \ \phi^+_2 (r)  \phi^+_2 (\tilde{r})
=\frac{1}{2l+1}(a+r)^{-l-1}(a+\tilde{r})^l
\nonumber\\
&&-\frac{a^{2l+1}\Big(a^2-2l(l+1)(4q_1+q_2)\Big) }{2(2l+1)(l+1)\Big(a^2-l(4q_1+q_2)\Big)}(a+r)^{-l-1}(a+\tilde{r})^{-l-1}.
\end{eqnarray}
Then, the potential (\ref{potential}) can be written as
\begin{eqnarray} \label{potential-1}
A_t(x, \tilde{x} ) &=& - e \sum_{l=0}^\infty
\left(2l+1\right) P_l(\cos\gamma) g_l(r,\tilde r) \nonumber\\
&=&-\frac{e}{a+r}\sum_{l=0}^\infty
\Big(\frac{a+\tilde{r}}{a+r}\Big)^l  P_l(\cos\gamma)
\nonumber\\
&&+\frac{e}{2}\sum_{l=0}^\infty
 P_l(\cos\gamma)\frac{a^{2l+1}\Big(a^2-2l(l+1)(4q_1+q_2)\Big) }{(l+1)\Big(a^2-l(4q_1+q_2)\Big)}(a+r)^{-l-1}(a+\tilde{r})^{-l-1}.
\end{eqnarray}
The first term of the equation (\ref{potential-1}) can be easily calculated using a series expansion
\begin{eqnarray}
\sum\limits_{l=0}^\infty t^l P_l(x) = \frac{1}{\sqrt{1-2xt+t^2}},
\end{eqnarray}
\begin{eqnarray}
A_t^M(x, \tilde{x} )&=&-\frac{e}{a+r}\sum_{l=0}^\infty
\Big(\frac{a+\tilde{r}}{a+r}\Big)^l  P_l(\cos\gamma) \nonumber \\ &=&-\frac{e}{\sqrt{(a+r)^2-2(a+r)(a+\tilde{r})\cos(\gamma)+(a+\tilde{r})^2}}.
\end{eqnarray}
 This expression diverges at $r\longrightarrow \tilde{r}$, and must be renormalized.

\section{The process of renormalization and the result}
The procedure for determining the self-force requires renormalization of
vector potential $A_k (x; \tilde x)$, which diverges at the limit $x \rightarrow \tilde x$ (see, for example, works \cite{1Roth:2004, 2Roth:2004}).
Renormalization is achieved by subtracting the DeWitt-Schwinger counter-term $A_{\mbox{\tiny \sl DS}}(x; \tilde x)$ from $A_t(x; \tilde x)$, and then using the approximation $x \rightarrow \tilde x$
\begin{equation} \label{ren}
A_{ren}(x)=\lim_{\tilde x \rightarrow x}\left( A_t(x; \tilde x)
-A_{\mbox{\tiny DS}}(x; \tilde x) \right).
\end{equation}
For a scalar charge at rest in a static, curved space-time, the DeWitt-Schwinger counter-term $A_{\mbox{\tiny \sl DS}}(x; \tilde x)$, which must be subtracted, has the following form \cite{Khu:10}
        \begin{equation} \label{phiDS}
        A_{\mbox{\tiny DS}}(x^i; \tilde x^i)=
        -e \left(\frac{1}{\sqrt{2 \sigma}}
        +\frac{\partial g_{t t}(\tilde x)}{ \partial \tilde x^i}
        \frac{\sigma^{{i}}}{4g_{t t}(\tilde x)\sqrt{2 \sigma}}
        \right),
        \end{equation}
where \cite{Synge,Popov:2007}
\begin{eqnarray}
{\sigma^i}&=&-\left(x^i-\tilde x^i\right)
        -\frac12 \Gamma^{i}_{{j}{k}}\left(x^j-{\tilde x^j}\right)\left(x^k-{\tilde x^k}\right)
        \nonumber \\ &&
        -\frac16 \left( \Gamma^{i}_{{j}{m}} \Gamma^{m}_{{k}{l}}
        +\frac{\partial \Gamma^{i}_{{j}{k}}}{\partial {\tilde x^l}}\right)
        \left(x^j-{\tilde x^j}\right)\left(x^k-{\tilde x^k}\right)\left(x^l-{\tilde x^l}\right)
        +O\left(\left(x-{\tilde x}\right)^4\right), \nonumber \\
\sigma &=& \frac{g_{i j }(\tilde x)}{2}  {\sigma^i} {\sigma^j},
\end{eqnarray}
$\Gamma^{i}_{{j}{k}}$  -the Christoffel symbols, calculated at $\tilde x$.
DeWitt-Schwinger counterterm $A_{\mbox{\tiny \sl DS}}(x; \tilde x)$ in the limit $t=\tilde t,\theta= \tilde \theta, \varphi=\tilde \varphi$ can be calculated easily using the metric (\ref{metric})
        \begin{equation} \label{phiDSf}
        A_{\mbox{\tiny DS}}(r; \tilde r)=-\frac{e}{|r-\tilde r|}.
        \end{equation}
Thus, we get an expression for the renormalized potential in the domain $r > 0$
\begin{eqnarray} \label{fin}
A_t^{ren}(r)
&=& \lim_{\tilde r \rightarrow r} \left(A_t(r, \tilde{r} ) - A_{\mbox{\tiny DS}}(r, \tilde{r})\right)
\nonumber\\
&=&\frac{e}{2}\sum_{l=0}^\infty
 \frac{a^{2l+1}\Big(a^2-2l(l+1)(4q_1+q_2)\Big) }{(l+1)\Big(a^2-l(4q_1+q_2)\Big)}(a+r)^{-2l-2}.
\end{eqnarray}
$A_{ren}$ matches this expression in the region $r < 0,$ due to the symmetry of the problem.
The potential of self-interaction and the tetrad component of the self-force have the form
       \begin{equation} \label{USf}
        U^{self}=-\frac{e}{2}A_t^{ren},
        \end{equation}
      \begin{equation} \label{Sf}
       F^{(r)}=-\frac{\partial U^{self}}{\partial r}
       =-\frac{e^2}{2} \sum_{l=0}^\infty  \frac{a^{2l+1}\Big(a^2-2l(l+1)(4q_1+q_2)\Big) }{\Big(a^2-l(4q_1+q_2)\Big)}(a+r)^{-2l-3}
        \end{equation}

\includegraphics[width=14cm]{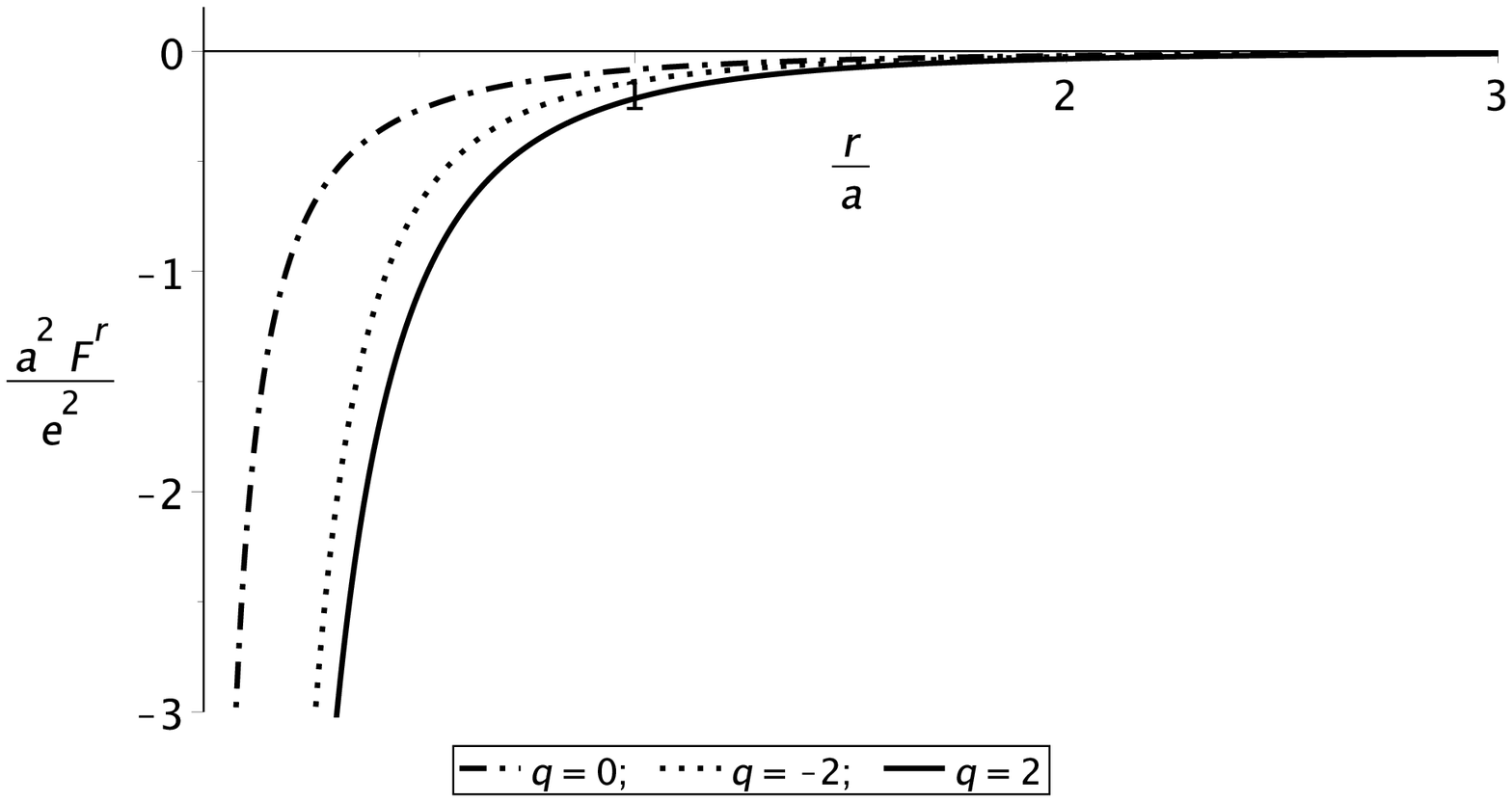}

\section{Conclusion}

The divergence of $F^{(r)}$ in the vicinity of $r=0$ is related to the disadvantage of the considered wormhole model. Its usage in this region is incorrect. Numerous descriptions of the effect of self-interaction in smooth-throated wormholes show that such divergence does not occur in such wormholes.

In the special case $4 q_1 +q_2 =0$ the result \eqref{Sf} coincides with the one discussed earlier in \cite{KhusBakh:2007}.

\section{Acknowledgements}

The work of A.A.P was partly funded by the development program of the Regional Scientific and Educational Mathematical Center of the Volga Federal District, agreement N 75-02-2022-882.

\bibliography{aap.bib}

\end{document}